\documentclass[journal]{IEEEtran}
\usepackage{amsmath}
\usepackage[ruled]{algorithm2e}
\usepackage{algorithmic}
\usepackage{bm}
\usepackage{tabulary}
\usepackage{amssymb}
\usepackage{graphicx}
\usepackage{subfigure}

\ifCLASSINFOpdf
\else
\fi
\begin{document}
%
\title{Bayesian Physics-Informed Neural Networks \\for the Subsurface Tomography \\based on the Eikonal Equation}
%
%
%

\author{Rongxi~Gou,
        Yijie~Zhang,
        Xueyu~Zhu, and
Jinghuai Gao, \IEEEmembership{Member,~IEEE}
\thanks{Manuscript received November 16, 2022. R. Gou and Y. Zhang would like to thank National Natural Science Foundation of China (42174137). X. Zhu was supported by the Simons Foundation (504054). The work of J. Gao is supported by National Key R\&D Program of China (2020YFA0713400). (\textit{Corresponding authors: Xueyu Zhu and Yijie Zhang}).

R. Gou, Y. Zhang and J. Gao are with the School of Information and Communications Engineering, Xi'an Jiaotong University, Xi'an, Shaanxi 710049, China (e-mail: grx123456@stu.xjtu.edu.cn; zhangyijie2016@mail.xjtu.edu.cn; jhgao@mail.xjtu.edu.cn).

X. Zhu is with the Department of Mathematics, The University of Iowa, Iowa City, IA 52246 (e-mail: xueyu-zhu@uiowa.edu).}}

%
%

\markboth{IEEE TRANSACTIONS ON GEOSCIENCE AND REMOTE SENSING}%
{Shell \MakeLowercase{\textit{et al.}}: Bare Demo of IEEEtran.cls for IEEE Journals}
%



\maketitle

\begin{abstract}
The high cost of acquiring a sufficient amount of seismic data for training has limited the use of machine learning in seismic tomography. In addition, the inversion uncertainty due to the noisy data and  data scarcity is less discussed in conventional seismic tomography literature. To mitigate the uncertainty effects and quantify their impacts in the prediction, the so-called Bayesian Physics-Informed Neural Networks (BPINNs) based on the eikonal equation are adopted to infer the velocity field and reconstruct the travel-time field. In BPINNs, two inference algorithms including Stein Variational Gradient Descent (SVGD) and Gaussian variational inference (VI) are investigated for the inference task. The numerical results of several benchmark problems demonstrate that the velocity field can be estimated accurately and the travel-time can be well approximated with reasonable uncertainty estimates by BPINNs. This suggests that the inferred velocity model provided by BPINNs may serve as a valid initial model for seismic inversion and migration.
\end{abstract}


%
\IEEEpeerreviewmaketitle

\section{Introduction}
%
%
%
%
\IEEEPARstart{S}{eismic} tomography is one of the most popular methods for  studying the structure of the underground. Particularly, travel-time tomography is an effective and mature technique to invert subsurface structure based on the ray theory \cite{bording1987applications}. The widely used partial differential equation (PDE) in travel-time tomography is the eikonal equation, which is the high-frequency approximation for the wave equation \cite{2002Quantitative}. The eikonal equation can be solved through several numerical methods, including the finite difference method \cite{vidale1990finite}, the fast marching method \cite{rawlinson2004wave} and the fast sweeping method \cite{zhao2005fast}. Travel-time tomography \cite{hole1992nonlinear, leung2006adjoint,9908560} can be typically solved by minimizing the misfit of the observed travel-time and simulated travel-time based on the eikonal equation with a synthetic velocity model, 
which has been applied to image the source area of the earthquake \cite{2020Eikonal} and guide the full waveform inversion \cite{treister2017full}. 
Nonetheless, traditional methods for travel-time tomography face several challenges: they often require a good initial model, which can affect the approximation quality significantly. In addition, since the measured data always contain noises, it is essential to quantify its impacts on the inversion results, which is useful to interpret the estimated velocity model.  However, this is a lack of study in the traditional travel-time tomography literature \cite{hole1992nonlinear, leung2006adjoint, 9908560, 2020Eikonal}. 
Therefore, it is desirable to develop flexible algorithms that are not only less sensitive to the initial velocity model and noises but also provide uncertainty estimates on the estimated velocity model. 

Recently, there is a growing interest to leverage machine learning power to solve the inverse problems based on the PDEs in order to overcome the shortcomings of the traditional methods \cite{Araya2017Deep, yang2019deep}. 
However, the performance of purely data-driven machine learning approaches is heavily dependent on the quality of training data, which could lead to poor performance for sparse/noisy datasets \cite{bianco2019high, yildirim2022machine}. 
Many recent efforts in the scientific machine learning community \cite{baker2019workshop, raissi2019physics, lu2021deepxde, psaros2022uncertainty} have been focused on these challenges. Notably, Physics-Informed Neural Networks (PINNs) have been demonstrated to solve a variety of problems with small datasets \cite{raissi2019physics}, where the underlying governing equations are introduced as a regularization term into the loss function. For seismic applications, PINNs have been used to construct the travel-time field from sparse observed travel-time data \cite{smith2020eikonet,bin2021pinneik} and solve the corresponding inverse problems to 
infer the velocity field \cite{waheed2021pinntomo} based on the eikonal equation. In addition, PINNs are also used for the full waveform inversion based on acoustic wave equation \cite{rasht2021physics}. Besides, PINNs have also been used in electrical properties tomography \cite{9814346} and cardiac electrophysiology \cite{2020Physics}.

Recently, Bayesian approaches have been integrated into PINNs framework to better account for the uncertainty and  provide the uncertainty estimations of predicted parameters by the network. Notably, the so-called Bayesian Physics-Informed Neural Networks (BPINNs) use physics equations as prior knowledge to compensate for the lack of training data while  Bayesian inference is used to predict the uncertainty of output results. BPINNs are applied for the fluid flow reconstruction based on the Navier-Stokes equation, where the stein variational gradient descent (SVGD) algorithm is used to enable the efficient Bayesian inference \cite{sun2020physics}. For the cardiac electrophysiology, the BPINNs based on the eikonal equation is used to estimate velocity fields \cite{ceccarelli2021bayesian}. In the context of geophysics, the hypocentre inversion is investigated by BPINNs with Stein variational inference to handle highly multimodal posterior distributions efficiently \cite{smith2022hyposvi}.  BPINNs with Laplacian approximation are used for hypocentre estimation and show promising results in estimating the locations of the hypocentre and providing uncertainty estimate \cite{2022uqpinnhypo}. 

To address the high dimensional issues in Bayesian inference, Variational Inference (VI) and Markov chain Monte Carlo (MCMC) method have been proposed. In contrast with  the traditional MCMC method, VI is a deterministic method approximating the target distribution by minimizing Kullback-Leibler (KL) divergence, which 
is computationally efficient for large datasets in general. Particularly, Blundell proposed a VI algorithm called Gaussian Variational Inference for neural networks \cite{blundell2015weight}. 
However, the accuracy of VI depends on the set of pre-defined distributions to approximate the target distribution. To address this problem, SVGD, a general variational inference algorithm that uses a set of particles rather than distributions to approximate the target distribution \cite{liu2016stein}, has been demonstrated to be efficient in different applications 
\cite{ceccarelli2021bayesian,smith2022hyposvi,sun2020physics}.

As previously stated, the traditional approaches for travel-time tomography often require an initial model and do not offer uncertainty estimates about the inferred velocity model. 
To address these issues,  motivated by the above developments in the scientific machine-learning community, we present a Bayesian physics-informed machine-learning framework for travel-time tomography with a limited number of observed data. Specifically, we use neural networks to approximate the travel-time fields and velocity field models. In addition to the data misfit, we also incorporate the ekional equation into the loss to acknowledge the underlying governing physics. To cope with uncertainty, we formulate the problem in the Bayesian framework and investigate the inference performance of the BPINNs using two popular methods from the literature -  Gaussian Variational Inference and SVGD.  
The contributions of this work include the following:
\begin{itemize}
    \item We adopt BPINNs for travel-time tomography, given a limited amount of the observed travel-time from the surface and the wells. To further improve the prediction performance, the velocity data from the specific locations are provided. 
    \item We introduced  depth-dependent velocity uncertainty  to better account for the data uncertainty of subsurface tomography.
    \item We demonstrated that BPINNs can provide a reasonably accurate  velocity field and meaningful uncertainty estimate for both the velocity field and travel-time field. 
    \item We demonstrated that the velocity model parameterized by a neural network is randomly initialized and  does not require a good initial velocity model in contrast to the traditional methods.
\end{itemize}

This paper is organized as follows. In Section \ref{sec:setup}, we introduce the eikonal equation and travel time tomography setup. Then we briefly discuss PINNs, BPINNs with Gaussian Variational Inference, and SVGD.  The depth-dependent uncertainty is introduced to better account for the data uncertainty under the context of travel time tomography in Section \ref{sec:method}. Several numerical benchmark  problems are provided to demonstrate the effectiveness of the method in Section \ref{sec:result}. A complex velocity model is also discussed in Section \ref{sec:discussion}. Finally, we conclude in Section \ref{sec:conclusion}.

\section{Problem Setup}
\label{sec:setup}
The propagation of seismic waves through the underground obeys Fermat's principle \cite{moser1991shortest}. Fermat's principle states that the path by a ray between two given points prefers the one with the shortest travel-time. For example, the direct wave would travel in a straight line from the source to the receiver if the underground is isotropic and homogeneous. Due to the compositional layering and tectonic structure, the seismic waves would be reflected and refracted, where the wave shall travel along the path with the shortest travel-time. 
In the literature, the {\it eikonal} equation \eqref{eq:eikonal} is used to model the relationship between travel-time and velocity field as follows: 
\begin{equation}
\label{eq:eikonal}
    \left\{
    \begin{aligned}
    {\lvert\nabla{T(\textbf{x$_s$},\textbf{x})}\rvert}^{2} &= \frac{1}{v^{2}(\textbf{x})},\qquad\forall{x}\in\Omega,\\
    T(\textbf{x$_s$},\textbf{x$_s$}) &= 0,
    \end{aligned}
    \right.
\end{equation}
where $T(\textbf{x$_s$},\textbf{x})$ represents the travel-time from the location of the point-source $\textbf{x$_s$}$ to any point $\textbf{x}$ in the domain $\Omega$, $v(\textbf{x})$ is the velocity defined in $\Omega$. 

Since the singular point exits at the point-source in equation \eqref{eq:eikonal}, 
we factorize $T(\textbf{x$_s$},\textbf{x})$ into two factors \cite{waheed2021pinntomo, bin2021pinneik} as follows:
\begin{equation}
\label{eq:factorize}
\begin{aligned}
{T(\textbf{x$_s$},\textbf{x})} = T{_0}(\textbf{x$_s$},\textbf{x})\tau(\textbf{x$_s$},\textbf{x}),
\end{aligned}
\end{equation}
where
\begin{equation}
\label{eq:factorize-2}
\enspace{T_0(\textbf{x$_s$},\textbf{x})} = \frac{\lvert{\textbf{x}-\textbf{x$_s$}\rvert}}{v(\textbf{x$_s$})}.
\end{equation}
Substituting equation \eqref{eq:factorize} into equation \eqref{eq:eikonal}, we get  the residual $\mathcal{R}(\textbf{x}_{s},\textbf{x})$  of the factorized eikonal equation: 

\begin{equation}
\label{eq:factorized eikonal}
    \mathcal{R}(\textbf{x}_{s},\textbf{x})=\left\{
    \begin{aligned}
    {\lvert{\nabla{(T_0(\textbf{x$_s$},\textbf{x})}\tau(\textbf{x$_s$},\textbf{x}))}\rvert}^2-
    \frac{1}{v^{2}(\textbf{x})}&= 0,\qquad\forall{\textbf{x}}\in\Omega,\\
    \tau(\textbf{x$_s$},\textbf{x}_{s})-1 &= 0,
    \end{aligned}
    \right.
\end{equation}

In this work, our goal is to infer velocity field $v(\textbf{x})$ and reconstruct travel-time $T(\textbf{x$_s$},\textbf{x})$ with a limited number of datasets.
The measured velocity data $v(\textbf{x})$ could be acquired from the well-logs \cite{pirson1963handbook}, which is rescaled as follows:
\begin{equation}
\begin{aligned}
v(\textbf{x}) = \frac{{v}(\textbf{x})-v_{\min}}{v_{\max}-v_{\min}},
\end{aligned}
\label{eq:normalize}
\end{equation}
$v_{\max}$ and $v_{\min}$ represent the chosen scaling factors. 

\section{Methods}
\label{sec:method}
In this section, we shall briefly introduce PINNs  and their application for ekinoal equations, then discuss Bayesian neural networks and BPINNs for ekinoal equations.
\subsection{Physics informed neural networks}

Standard PINNs  approximate the unknown solution $y(\textbf{x})$ of the underlying PDE by a neural network $\mathcal{N}(\textbf{x};\bm{\theta})$ parameterized  by $\bm{\theta}$. In contrast with  purely data-driven machine learning algorithms, PINNs incorporate the residual of the underlying governing  equation into the loss function in order to provide additional knowledge. The loss function for PINNs can be written as follows:
\begin{equation}
\label{eq:genpc loss}
\begin{split}
\mathcal{L}(\bm{\theta}) = &\underbrace{\frac{\lambda_1}{|\mathcal{T}_{f}|}\sum_{\textbf{x}\in\mathcal{T}_{f} }||\mathcal{F}(\mathcal{N}(\textbf{x};\bm{\theta}),\textbf{x})||}_\text{model driven}+\\&\underbrace{\frac{\lambda_2}{|\mathcal{T}_{d}|}\sum_{\textbf{x}\in\mathcal{T}_{d}}||\mathcal{N}(\textbf{x};\bm{\theta})-y(\textbf{x})||}_\text{data driven},
\end{split}
\end{equation}
where $\lambda_1$ and $\lambda_2$ are the weights for each term. $\mathcal{F}$ is the residual of the governing physics equation.  $\mathcal{T}_{f}$ denotes the  set to the locations to evaluate the residual  $\mathcal{F}$, $|\mathcal{T}_{f}|$ is the number of data points. $\mathcal{T}_{d}$ represents the locations of the observed data set, while $|\mathcal{T}_{d}|$ is the number of observed data.

For seismic tomography in this work, two independent  neural networks $\mathcal{N}_{\tau}$ and $\mathcal{N}_{v}$ are adopted to approximate the travel-time factor and velocity field:
\begin{equation}
\begin{aligned}
\hat{\tau}(\textbf{x}_s,\textbf{x})&=\mathcal{N}_{\tau}(\textbf{x}_s,\textbf{x};\bm{\theta}_{\tau}),\quad 
\hat{v}(\textbf{x})&=\mathcal{N}_{v}(\textbf{x};\bm{\theta}_{v}),
\end{aligned}
\label{eq:NN output}
\end{equation}
where $\bm{\theta}_\tau$ and $\bm{\theta}_v$ are the weights and bias of the neural networks $\mathcal{N}_{\tau}$ and $\mathcal{N}_{v}$, respectively. Here, the sigmoid function is used in the last layer of $\mathcal{N}_{v}$ to restrict the output of $\mathcal{N}_{v}$ between 0 and 1.  
Subject to the eikonal equation \eqref{eq:factorized eikonal}, 
the corresponding loss function can be formulated as follows:
\begin{equation}
\label{eq:constrained loss}
\begin{aligned}
\mathcal{L}(\bm{\theta}) = &\frac{\lambda_1}{|\mathcal{T}_{r}|}\sum_{(\textbf{x}_{s},\textbf{x})\in\mathcal{T}_{r}}||\mathcal{R}(\textbf{x}_{s},\textbf{x})||+\frac{\lambda_2}{|\mathcal{T}_{v}|}\sum_{\textbf{x}\in\mathcal{T}_{v}}||\hat{v}(\textbf{x})-v(\textbf{x})||\\&+\frac{\lambda_3}{|\mathcal{T}_{\tau}|}\sum_{(\textbf{x}_{s},\textbf{x})\in\mathcal{T}_{\tau}}||\hat{\tau}(\textbf{x}_s,\textbf{x})-{\tau}(\textbf{x}_s,\textbf{x})||,
\end{aligned}
\end{equation}
where $ \bm{\theta}=[\bm{\theta}_\tau, \bm{\theta}_v]$. $\lambda_1$, $\lambda_2$ and $\lambda_3$ are the weights for each term. $\mathcal{T}_{r}$ denotes the set of the locations of the source-receiver pairs to evaluate  the residual $\mathcal{R}(\textbf{x}_{s},\textbf{x})$ in equation \eqref{eq:factorized eikonal}. $\mathcal{T}_{v}$ and $\mathcal{T}_{\tau}$ are the sets  of the locations of the observed velocity and travel-time, respectively. The parameters $\bm{\theta}$ is typically optimized by ADAM \cite{kingma2014adam}. Once the parameters are optimized, the approximated travel-time factor and velocity field are obtained.

\subsection{BPINNs for the eikonal equation}
\label{bnn-ek}
Nonetheless, vanilla PINNs \cite{raissi2019physics} have limited capability to account for the uncertainty from the model, unknown parameters, and noisy data. To better quantify their impacts on the neural network outputs, we shall first introduce the Bayesian Neural Network (BNN), followed by its adaption in the context of PINNs. 
After that, we shall discuss two variants of BPINNs for inference. 

\subsubsection{Bayesian Neural Network}
\label{bnn}
For traditional neural networks, the network weights and bias $\bm{\theta}$ are assumed to be deterministic 
values \cite{2020Bayesian}. 
In contrast, BNNs consider $\bm{\theta}$ as random variables with specific distributions, that can be learned based on the Bayesian theorem: 
\begin{equation}
        p(\bm{\theta}|\mathcal{D})=\frac{p(\mathcal{D}|\bm{\theta})p(\bm{\theta})}{p(\mathcal{D})}\sim
        p(\mathcal{D}|\bm{\theta})p(\bm{\theta}),
        \label{eq:Bayes' theorem}
\end{equation}
where $\bm{\theta}$ and $\mathcal{D}$ represent the network parameters and measurements, respectively.
By using \eqref{eq:Bayes' theorem}, the posterior distribution of the parameters $p(\bm{\theta}|\mathcal{D})$ can be computed by the prior distribution $p(\bm{\theta})$ and the likelihood $p(\mathcal{D}|\bm{\theta})$. Since the dataset is independent of $\bm{\theta}$, $p(\mathcal{D})$ can be treated as a normalized constant during the training process. Once the network is trained for the specific dataset $\mathcal{D}$, the predicted results and the associated uncertainty can be approximated by the posterior sample mean value and standard deviation:
\begin{equation}
    \begin{aligned}
    \mathbb{E}_{p(\bm{\theta}|\mathcal{D})}[\textbf{y}|\textbf{x},\mathcal{D}] &\approx \frac{1}{M}\sum_{i=1}^{M}\mathcal{N}(\textbf{x};\bm{\theta}_{i}),\\
    {\rm{Var}}_{p(\bm{\theta}|\mathcal{D})}[\textbf{y}|\textbf{x},\mathcal{D}] &\approx \frac{1}{M}\sum_{i=1}^{M}(\mathcal{N}(\textbf{x};\bm{\theta}_{i})-\mathbb{E}_{p(\bm{\theta}|\mathcal{D})}[\textbf{y}|\textbf{x},\mathcal{D}])^{2},
    \label{eq:BNNoutput}
    \end{aligned}
\end{equation}
where $\mathcal{N}(\textbf{x};\bm{\theta}_{i})$ 
represent the Bayesian neural network with  the corresponding parameter the $i$-th sample of  
$\bm{\theta}$, respectively. 
$M$ is the number of samples drawn from the posterior distribution of neural network parameters $\bm{\theta}$. $\textbf{x}$ and  $\textbf{y}$ are the input and output of the neural network. 

\subsubsection{Bayesian Physics-informed Neural Networks}
\begin{figure}
    \includegraphics[width=3.50in,height=3.00in]{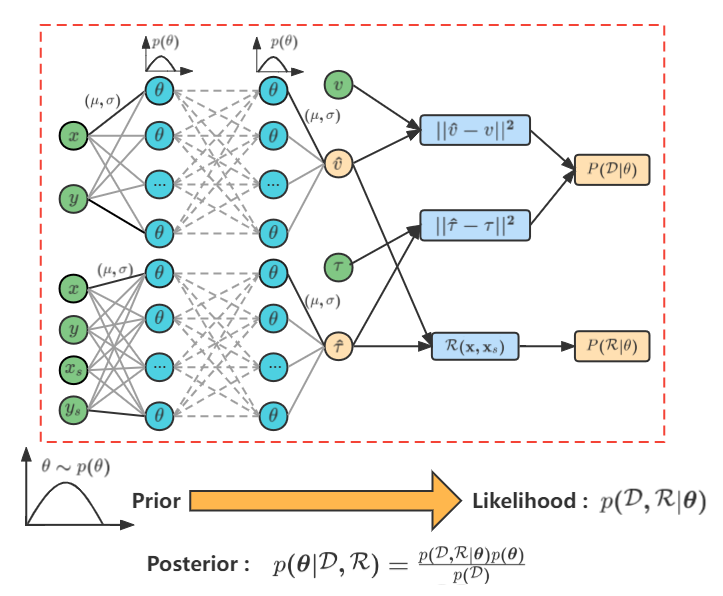}
	\caption{The structure of BPINNs for seismic tomography based on the eikonal equation. $p(\bm{\theta})$ is the prior distribution of network parameters, $p(\bm{\theta}|\mathcal{D})$ represents the likelihood of predicted velocity and travel-time. The posterior $p(\bm{\theta}|\mathcal{D},\mathcal{R})$ is  derived based on Bayesian theorem.}
	\label{fig:NN structure}
\end{figure}

Similar to traditional neural networks, BNNs may suffer from  a lack of data. Incorporating the underlying governing equation to BNN can provide additional prior knowledge to improve the generalization, which is referred to as {\it BPINNs} \cite{ceccarelli2021bayesian,2022uqpinnhypo,smith2022hyposvi,sun2020physics}. With the  physical constraints, Bayes' theorem under the context of BPINNs can be formulated as follows:
\begin{equation}
        p(\bm{\theta}|\mathcal{D},\mathcal{R})=\frac{p(\mathcal{D},\mathcal{R}|\bm{\theta})p(\bm{\theta})}{p(\mathcal{D})}\sim
        p(\mathcal{D},\mathcal{R}|\bm{\theta})p(\bm{\theta}),
        \label{eq:pc Bayes' theorem}
\end{equation}
where $\mathcal{R}$ represents the residual of the eikonal equation in \eqref{eq:factorized eikonal}. The likelihood $p(\mathcal{D},\mathcal{R}|\bm{\theta})$ is
\begin{equation}
    \log p(\mathcal{D},\mathcal{R}|\theta)=\log p(\mathcal{D}|\theta)+\log p(\mathcal{R}|\theta),
\label{eq:BPINNs}
\end{equation}
where $p(\mathcal{D}|\bm{\theta})$ is the likelihood of the observed data and $p(\mathcal{R}|\bm{\theta})$ represents the likelihood about the physics model. 
The whole structure of BPINNs for seismic tomography is illustrated in Figure \ref{fig:NN structure}. Two independent fully-connected neural networks $\mathcal{N}_{\tau}$ and $\mathcal{N}_{v}$ are adopted to approximate the velocity field and the travel-time field.

In many practical applications, computing posterior distribution is not tractable due to the high dimension of parameter space.
To address this issue, variational inference (VI) is proposed to approximate the target distribution by another reparameterized density function. In particular, Blundell \cite{blundell2015weight} proposed a Gaussian VI method with backpropagation for deep learning. However, the performance of VI is highly dependent on the choice of the approximated density function. Alternatively, Stein Variational Gradient Descent (SVGD) \cite{liu2016stein} is proposed to approximate the posterior distribution with a set of particles. In this work, we shall leverage and compare these two algorithms to perform the inference task for BPINNs. 

{\bf BPINNs with Gaussian Variational Inference}.
Computing the posterior distribution \eqref{eq:pc Bayes' theorem} can be intractable for high dimensional problems, which is common for neural networks. To address this issue, variational inference (VI) \cite{blei2017variational, graves2011practical} has been adopted to approximate $p(\bm{\theta}|\mathcal{D},\mathcal{R})$ with a predefined family of distributions $Q(\bm{\theta};\boldsymbol{\zeta})$, where $\boldsymbol{\zeta}=(\zeta_0,\zeta_1,\cdots,\zeta_n)$ parameterize the distribution of $Q(\bm{\theta};{\boldsymbol{\zeta}})$. In this work, 
the distribution family is chosen to be normal distributions, where $\boldsymbol{\zeta} = (\zeta_{\mu}, \zeta_{\sigma})$, $\zeta_{\mu}$ and $\zeta_{\sigma}$ represent the corresponding mean and standard deviation, respectively. Under the assumption that the parameters of the neural network are independent, $Q(\bm{\theta};\boldsymbol{\zeta})$ can be expressed as: 
\begin{equation}
    Q(\bm{\theta};\boldsymbol{\zeta})=\prod_{i=1}^{d_{\bm{\theta}}}q(\theta_{i};\zeta_{\mu,i}, \zeta_{\sigma,i}),
    \label{eq:VI Q}
\end{equation}
where $d_{\bm{\theta}}$ is the number of neural network’s parameters. $\bm{\theta}_{i}$ represents $i$-th parameter of the neural network, obeying one-dimensional Gaussian distribution $\mathcal{N}(\zeta_{\mu,i},2\log(1+e^{\zeta_{\sigma,i}}))$.
Instead of using the sampling approach, such as Markov chain Monte Carlo (MCMC), VI reformulates it as a deterministic optimization problem by minimizing the Kullback-Leibler (KL) divergence between the posterior distribution  $p(\bm{\theta}|\mathcal{D},\mathcal{R})$ and the proposed distribution $Q(\bm{\theta};{\boldsymbol{\zeta}})$:
\begin{equation}
\begin{split}
    &D_{\rm{KL}}(Q(\bm{\theta};\boldsymbol{\zeta})||p(\bm{\theta}|\mathcal{D},\mathcal{R}))= \int{Q(\bm{\theta};\boldsymbol{\zeta})\log\frac{Q(\bm{\theta};\boldsymbol{\zeta})}{p(\bm{\theta}|\mathcal{D},\mathcal{R})}d\bm{\theta}}\\&\propto\int{Q(\bm{\theta};\boldsymbol{\zeta})\log\frac{Q(\bm{\theta};\boldsymbol{\zeta})}{p(\bm{\theta})p(\mathcal{D},\mathcal{R}|\bm{\theta})}d\bm{\theta}}
    \approx \mathbb{E}_{\bm{\theta}\sim{Q}}[\log Q(\bm{\theta};\boldsymbol{\zeta})\\&- \log p(\bm{\theta})-\log p(\mathcal{D}|\bm{\theta})-\log p(\mathcal{R}|\bm{\theta})],
    \label{eq:KL}
\end{split}
\end{equation}
It can be shown that KL divergence has its minimum value when $Q(\bm{\theta};\boldsymbol{\zeta})$ and $p(\bm{\theta}|\mathcal{D},\mathcal{R})$ follow the same distribution, which enables us to approximate target distribution \cite{graves2011practical}.

To simplify computations, we assume that prior distribution $p(\bm{\theta})$ follows a Gaussian distribution   $\mathcal{N}(\mathbf{0}_{d_{\bm{\theta}}},\mathbf{I}_{d_{\bm{\theta}}})$.
The likelihoods of the data and the model are assumed to follow zero-mean Gaussian distributions:
\begin{equation}
\begin{aligned}
\log p(\mathcal{D}|\bm{\theta}) \propto (-\frac{1}{2{\sigma_{\mathcal{D}_v}^2}}\sum_{\textbf{x}\in\mathcal{T}_{v}}(\hat{v}({\mathbf{x}})-v({\mathbf{x}}))^2)+\\(-\frac{1}{2{\sigma_{\mathcal{D}_\tau}^2}}\sum_{(\textbf{x}_{s},\textbf{x})\in\mathcal{T}_{\tau}}(\hat{\tau}({\mathbf{x}_{s}},{\mathbf{x}})-\tau({\mathbf{x}_{s}},{\mathbf{x}}))^2),\\
\log p(\mathcal{R}|\bm{\theta})\propto-\frac{1}{2{\sigma^2_{\mathcal{R}}}}\sum_{(\textbf{x}_{s},\textbf{x})\in\mathcal{T}_{r}}(\mathcal{R}(\textbf{x}_{s},\textbf{x})-0)^2.
\end{aligned}
\label{eq:VI likelihood}
\end{equation}
where $\sigma_{\mathcal{D}_v}$ and $\sigma_{\mathcal{D}_\tau}$ represent the standard deviation for data likelihood $p(\mathcal{D}|\bm{\theta})$ of predicted velocity and travel-time factor. $\sigma_{\mathcal{R}}$ is the standard deviation for model likelihood $p(\mathcal{R}|\bm{\theta})$.

The minimization of KL divergence as \eqref{eq:KL} is performed through gradient backpropagation by Adam optimizer. The algorithm of variational inference is shown in Algorithm \ref{alg:VI}. 

\begin{algorithm}
    \label{alg:VI}
    \caption{BPINNs with Gaussian Variational Inference for seismic tomography}  
    
    \KwIn{A set of sparse, noisy travel-time and velocity data.} 
    
    \KwOut{Trained parameters $\bm{\zeta}$ through sampling the networks parameters $\bm{\theta}$.}
    
    \textbf{Initialization}: initialize $\bm{\zeta}=(\boldsymbol{\zeta_{\mu}},\bm{\zeta_{\sigma}})=\mathcal{N}(\mathbf{0}_{d_{\bm{\theta}}},\mathbf{I}_{d_{\bm{\theta}}})$, $d_{\bm{\theta}}$ is the number of neural network's parameters.
    
    \For {i=1:$N_e$}{
        \begin{enumerate}
            \item Sample $\bm{\epsilon}_j$ from $\mathcal{N}(\mathbf{0}_{d_{\bm{\theta}}},\mathbf{I}_{d_{\bm{\theta}}})$ independently, where $j=0,1,...,n$, $n$ is the number of \\sample  of networks parameters; 
            \item $\bm{\theta}_j = \bm{\zeta}_{\mu}+\log(1+\exp(\bm{\zeta}_{\sigma}))\odot\bm{\epsilon}_j$;
            \item $\mathcal{L}(\bm{\zeta}) = \frac{1}{n}\sum_{j=1}^{n} [\log Q(\theta_j;\bm{\zeta})-\log p(\bm{\theta}_j)-\log p(\mathcal{D}|\bm{\theta}_j)-\log p(\mathcal{R}|\boldsymbol{\theta}_j)]$; 
            \item Using Adam optimizer to upgrade $\bm{\zeta}$ based on the gradient of $\mathcal{L}(\bm{\zeta})$.
        \end{enumerate}
        }
\end{algorithm}

{\bf BPINNs with Stein Variational Gradient Descent}.
Alternatively, Stein Variational Gradient Descent (SVGD) is a general variational inference algorithm  based on Stein's identity and kernelized Stein discrepancy. SVGD employs a set of particles to approximate  the target posterior distribution by adjusting the distribution of the particles. 
To update these particles, Kernelized Stein discrepancy is adopted to find the steepest descent for the KL divergence \cite{liu2016stein}.

Similar to KL divergence, Stein's identity $\mathbb{E}_{\theta\sim q}[\mathcal{A}_{p}\rm{f}(\bm{\theta})]$  can measure the distance between two distributions, i.e. approximated distribution $q(\bm{\theta})$ and the target distribution $p(\bm{\theta})$ , here
\begin{equation}
    \mathcal{A}_{p}\rm{f}(\bm{\theta})=\nabla_{\bm{\theta}}\log {p(\bm{\theta})}\rm{f}(\bm{\theta})^{T}+\nabla_{\bm{\theta}}\rm{f}(\bm{\theta}),
    \label{eq:Stein's operator}
\end{equation}
where $\nabla$ denotes the gradient operator. 
$\rm{f}(\bm{\theta})$ is a smooth function and satisfied $\int_{-\infty}^{\infty}\nabla_{\bm{\theta}}(\rm{f}(\bm{\theta})q(\bm{\theta}))d\bm{\theta}=0$. When $q(\bm{\theta})$ and $p(\bm{\theta})$ follow the same distribution, Stein's identity $\mathbb{E}_{\bm{\theta}\sim q}[\mathcal{A}_{p}\rm{f}(\bm{\theta})]$ equals to zero. During the iteration, the parameters $\theta$ will be updated as follows: $\bm{\theta} = \bm{\theta}+\epsilon\phi(\bm{\theta})$, where $\epsilon$ is the step size and $\phi(\bm{\theta})$ represents the updating direction of parameters. According to Liu's work \cite{liu2016stein}, KL divergence is proved to decay fastest when

\begin{equation}
    \nabla_{{\epsilon}}\rm{KL}(q||p)= -\mathbb{E}_{\theta\sim{q}}[\rm{tr}(\mathcal{A}_{p}f(\bm{\theta}))].
    \label{eq:KL descent}
\end{equation}

Unlike Gaussian Variational Inference introduced in the previous subsection, SVGD employs a set of deterministic particles $\{\theta_{j}\}_{j=1}^{n}$ to minimize KL divergence instead of sampling from the variational distribution family. With the gradient direction, we can use the SVGD algorithm to update particles as follows: 
\begin{small}
\begin{equation}
\begin{split}
        \phi(\bm{\theta})&=\frac{1}{n}\sum_{j=1}^n[k(\bm{\theta}_{j}^t,\bm{\theta})\nabla_{\bm{\theta}_{j}^t}(\text{log} p(\bm{\theta}_{j}^t)+\text{log}p(\text{likelihood}))\\&+\nabla_{\bm{\theta}_{j}^t}k(\bm{\theta}_{j}^{t},\bm{\theta})],\\ \bm{\theta}_{j}^{t+1}&=\bm{\theta}_{j}^t+\epsilon_{t}\phi(\bm{\theta}_{j}^{t}),
        \label{eq:upgrade}
\end{split}
\end{equation} 
\end{small}
where $t$ represents $t$-th iteration, $k(\cdot,\cdot)$ represents a positive kernel function. In this work, radial basis function (RBF) 
is used,
\begin{equation}
\label{eq:RBF}
k(\bm{\theta}_{j}^t,\bm{\theta})=\exp(-\frac{{||\bm{\theta}-\bm{\theta}_{j}^t||}^2}{2l^2}),
\end{equation}
where $l$ is the median distance between the particles $\{\bm{\theta}_{j}\}_{j=1}^{n}$ to control the lengthscale of the kernel.

To simplify the expression of the distributions, normal distributions are selected for data likelihood $p(\mathcal{D}|\bm{\theta},\Sigma_{\mathcal{D}})$ and model likelihood $p(\mathcal{R}|\bm{\theta},\Sigma_{\mathcal{R}})$. The logarithmic likelihood function can be written as
\begin{equation}
\begin{aligned}
        &\log(\text{likelihood})=\log p(\mathcal{D}|\bm{\theta},\Sigma_{\mathcal{D}})+\log p(\mathcal{R}|\bm{\theta},\Sigma_{\mathcal{R}})\\&=\sum_{j=1}^{n}(\log p(\mathcal{D}|\bm{\theta}_{j},\sigma_{\mathcal{D},j})+\log p(\mathcal{R}|\bm{\theta}_{j},\sigma_{\mathcal{R},j})),
        \label{lh_svgd}
\end{aligned}
\end{equation}        
where $\bm{\theta}_{j}$ is the network parameters of $j$-th particle, $\bm{\theta}$ represents the network parameters of all $n$ particles. The diagonal matrices $\Sigma_{\mathcal{D}}=[\rm{diag}({\sigma^2_{\mathcal{D}_v}})_{n\times n},\rm{diag}(\sigma_{\mathcal{D}_\tau}^2)_{n\times n}]$ and $\Sigma_{\mathcal{R}}=\rm{diag}(\sigma_{\mathcal{R}}^2)_{n\times n}$ are the trainable parameters. $\Sigma_{\mathcal{D}}$ is the covariance matrix of the distributions $p(\mathcal{D}|\bm{\theta},\Sigma_{\mathcal{D}})$, $\Sigma_{\mathcal{R}}$ reflects the confidence about the physics model. $p(\mathcal{D}|\bm{\theta},\Sigma_{\mathcal{D}})$ and $p(\mathcal{R}|\bm{\theta},\Sigma_{\mathcal{R}})$ denotes the sum of data and model likelihood for each particle, respectively. Here, the likelihood for one particle can be represented as follows:
\begin{equation}
\begin{split}
        &\log p(\mathcal{D}|\bm{\theta}_{j},\sigma_{\mathcal{D},j}) \propto (-\frac{1}{2{\sigma^2_{\mathcal{D}_{v,j}}}}\sum_{\textbf{x}\in\mathcal{T}_{v}}(\hat{v}({\mathbf{x}})-v({\mathbf{x}}))^2+\\&{|\mathcal{T}_{v}|}\log(\frac{1}{\sigma_{\mathcal{D}_{v,j}}})) +(-\frac{1}{2{\sigma^2_{\mathcal{D}_{\tau,j}}}}\sum_{(\textbf{x}_{s},\textbf{x})\in\mathcal{T}_{\tau}}(\hat{\tau}({\mathbf{x}_{s}},{\mathbf{x}})-\tau({\mathbf{x}_{s}},{\mathbf{x}}))^2\\&+{|\mathcal{T}_{\tau}|}\log(\frac{1}{\sigma_{\mathcal{D}_{\tau,j}}})),\\
        &\log p(\mathcal{R}|\bm{\theta}_{j},\sigma_{\mathcal{R},j})\propto(-\frac{1}{2{\sigma^2_{\mathcal{R},j}}}\sum_{(\textbf{x}_{s},\textbf{x})\in\mathcal{T}_{r}}(\mathcal{R}(\textbf{x}_{s},\textbf{x})-0)^2+\\&{|\mathcal{T}_{r}|}\log(\frac{1}{\sigma_{\mathcal{R},j}})),
\label{particle likelihood}
\end{split}
\end{equation}
where $\sigma_{\mathcal{D},j}=[\sigma_{\mathcal{D}_{v,j}},\sigma_{\mathcal{D}_{\tau,j}}]$, $\sigma_{\mathcal{R},j}$ represents the standard deviation of data uncertainty and model uncertainty for particle $\bm{\theta}_{j}$. $\hat{v}$ and $\hat{\tau}$ in equation \eqref{particle likelihood} are the outputs from particle $\bm{\theta}_{j}$, while $v$ and $\tau$ are the corresponding training data. The corresponding posterior variance  is defined as
\begin{small}
\begin{equation}
\begin{aligned}
    {\rm{Var}}_{p(\bm{\theta}|\mathcal{D})}[\hat{\textbf{y}}|\textbf{x},\mathcal{D}] &\approx {\underbrace{\frac{1}{n}\sum_{j=1}^{n}(\mathcal{N}(\textbf{x};{\bm{\theta}_{j}})-\mathbb{E}_{p(\bm{\theta}|\mathcal{D})}[\textbf{y}|\textbf{x},\mathcal{D}])^{2}}_\text{variance of particles}} \\&+ {\underbrace{\mathbb{E}_{p(\bm{\theta}|\mathcal{D})}[\Sigma_{\mathcal{D}}]}_\text{predicted data uncertainty}}
\end{aligned}
\label{eq:uncertainty svgd}
\end{equation}
\end{small}
where $\hat{\textbf{y}}$ represents the predicted velocity and travel-time fields. The variances from $n$ particles provide the uncertainty  estimate contributed by the posterior distribution of network parameters, while the predicted data uncertainty represents the uncertainty due to measurement noise or lack of observed data. 
According to \eqref{lh_svgd}, the posterior distribution could be estimated through a set of particles $\{\bm{\theta}_{j}\}_{j=1}^{n}$. To summarize, we list the SVGD algorithm in Algorithm \ref{alg:SVGD}.

\begin{algorithm}
    \label{alg:SVGD}
    \caption{BPINNs based on Stein Variational Gradient Descent for seismic tomography} 
    \KwIn{A set of sparse, noisy  travel-time and velocity data.}
    \KwOut{Trained networks parameters $\{\bm{\theta}_{j}\}_{j=1}^{n}$ in order to reconstruct travel-time and estimate velocity field in the computational domain. $\Sigma_{\mathcal{D}}$ and $\Sigma_{\mathcal{R}}$ are the data variance and model uncertainty, respectively. $n$ is the number of particles.}
    \textbf{Initialization}:\\ Sample initial values for $\bm{\theta}$ from Gaussian distribution $\mathcal{N}(\mathbf{0}_{d_{\bm{\theta}}},\mathbf{I}_{d_{\bm{\theta}}})$ with $n$ particles, $d_{\bm{\theta}}$ is the number of neural network parameters $\bm{\theta}$. \\Sample each parameter in $\Sigma_{\mathcal{D}}$ from Gamma distribution $Gamma(\Sigma_{\mathcal{D}}|\alpha_1,\beta_1)$, $\alpha_1$ and $\beta_1$ are the shape and rate parameters, respectively. Sample each parameter in $\Sigma_{\mathcal{R}}$ from $Gamma(\Sigma_{\mathcal{D}}|\alpha_2,\beta_2)$.\\
    \For{t=0:$t_{\rm{end}}$}{
        \begin{enumerate}
            \item Calculate $\mathcal{L}(\theta^{t})=\log p(\bm{\theta}^{t})+\log p(\Sigma_{\mathcal{D}})+\log p(\Sigma_{\mathcal{R}})+\log(\text{likelihood})$ 
            \item Calculate $\phi(\bm{\theta}_{j}^{t})=\frac{1}{n}\sum_{j=1}^n[k(\bm{\theta}_{j}^t,\bm{\theta}^{t})\nabla_{\bm{\theta}^{t}}\mathcal{L}(\bm{\theta}^{t})+\nabla_{\bm{\theta}_{j}^t}k(\bm{\theta}_{j}^t,\bm{\theta}^{t})]$
            \item Update networks parameters $\bm{\theta}_{j}^{t+1}=\bm{\theta}_{j}^t+\epsilon_{t}\phi(\bm{\theta}_{j}^{t})$ and $\Sigma_{\mathcal{D}}$ by stochastic gradient descent
        \end{enumerate}
        }
\end{algorithm}

{\bf Depth-dependent uncertainty}.
It is worth noting that the data uncertainty $\sigma_{\mathcal{D}_v}$ in \eqref{lh_svgd} is assumed to be the learnable constant across the entire domain. Nevertheless in surface tomography, the receivers are not evenly distributed, and the number of receivers near the surface is typically more than those in the layer far from the surface. As a result, higher uncertainty about the velocity is expected as the depth from the surface increases. Besides, in most situations, as the depth increases, strata pressure rises, which leads to higher wave propagation velocity. This may cause higher uncertainty at the deep layer due to the noise included in measurement data. Motivated by this observation, we assume the  uncertainty of predicted velocity $\Sigma_{\mathcal{D}_{v}}$ 
follows the linear relationship with the depth $z$ as follows:
\begin{equation}
    \Sigma_{\mathcal{D}_v}(z)=\Sigma_{\mathcal{D}_v}( z_{\min})+\frac{z-z_{\min}}{z_{\max}-z_{\min}}\left [\Sigma_{\mathcal{D}_v}(z_{\max})-\Sigma_{\mathcal{D}_v}(z_{\min})\right],
    \label{eq:depth uncertainty}
\end{equation}
where $z_{\min}$ and $z_{\max}$ are the minimum and maximum of the depth of the velocity model, respectively. $\Sigma_{\mathcal{D}_v}(z_{\min})$ and $\Sigma_{\mathcal{D}_v}(z_{\max})$ are the trainable parameters that represent the corresponding uncertainty of the velocity  at $z_{\min}$ and $z_{\max}$, respectively. 

\section{Numerical examples}
\label{sec:result}
In this section, we shall assess the feasibility and performance of the proposed algorithms via several benchmark problems. 
We first study a 1D homogeneous model with an analytic solution to verify the inference capability of the proposed methods. Then a 2D model with an ellipsoidal inclusion is used to further compare the  of VI and SVGD. Finally, depth-dependent uncertainty is employed  to test the effectiveness of SVGD with additional prior knowledge of  uncertainty distribution. 

In the numerical examples, we use the CPU toolkit named scikit-fmm \cite{scikit-fmm} to generate the travel-time data for the synthetic velocity models. 
The velocity data (from the ground truth model) at certain locations are provided to mimic the detailed records from well-logs in the practical setup. Besides, we corrupted the observed travel-time and velocity with  Gaussian noises as follows:
\begin{equation}
\begin{split}
        v_{d}&=v(1+\epsilon_v),\quad\epsilon_{v}\sim\mathcal{N}(0,\sigma^2_{v})\\
        \tau_{d}&=\tau(1+\epsilon_{\tau}),\quad\epsilon_{\tau}\sim\mathcal{N}(0,\sigma^2_{\tau}),
        \label{eq:noise}
\end{split}
\end{equation}
where $\tau_d$ and $v_d$ are the measured travel-time factor and velocity, respectively. $\epsilon_{v}$ and $\epsilon_{\tau}$ are sampled from the distribution $\mathcal{N}(0,\sigma^2_{v})$ and $\mathcal{N}(0,\sigma^2_{\tau})$ independently, here $\sigma_{v}$ and $\sigma_{\tau}$ represent the level of noise for velocity and travel-time factor.

To quantify the performance of the proposed method, we compute the correlation coefficient $\gamma$ and the absolute relative error (ARE):

\begin{align}
    \gamma &= \frac{\sum_{i=1}^{n}({\hat{y}^i}-\bar{\hat{y}})\sum_{i=1}^{n}(y^{i}-\bar{y})}{{\sqrt{\sum_{i=1}^{n}({\hat{y}^i}-\bar{\hat{y}})^2}}{\sqrt{\sum_{i=1}^{n}(y^{i}-\bar{y})^2}}},\label{eq:ARE2} \\
    {\rm{ARE}} &= \frac{1}{n}\sum_{i=1}^{n}\frac{\lvert{\hat{y}^i}-y^i\rvert}{\lvert y^i\rvert},
\label{eq:ARE}
\end{align}
where $n$ is the total number of test points. $\hat{y}$ is the output results of BPINNs, $y$ is the ground truth corresponding to $\hat{y}$. $\bar{\hat{y}}$ 
and $\bar{y}$ are the mean values of predicted $\hat{y}$ and the ground truth $y$, respectively. All neural network is trained by the Pytorch with NVDIA GeForce RTX 2080 Ti GPU.

\subsection{1D ekional equation of homogeneous model}
To test the effectiveness of BPINNs in quantifying uncertainty, we first consider the 1D ekional equation of a homogeneous model with a constant velocity $v$ as follows, motivated by \cite{ceccarelli2021bayesian}: 
\begin{equation}
\label{eq:example1}
    \left\{
    \begin{aligned}
    {\lvert{\frac{\partial{T(x)}}{\partial{x}}}\rvert}^{2} &= \frac{1}{v^{2}},\qquad\forall{x}\in[0,2],\\
    T(x_s) &= 0,
    \end{aligned}
    \right.
\end{equation}
where $x$ is the location of the receiver. The source location $x_s$ is 0 km. The ground truth velocity is set to be $v=2$ km/s. In this case, the analytical travel time is $T(x)= \dfrac{x}{v}$.

We generate a synthetic dataset of size $n_\tau$ noisy measurements $\mathcal{D}=\left\{x_i,T_d(x_{i})\right\}_{i=1}^{n_\tau}$ by corrupting the noise-free data generated by the analytical travel-time with $5\%$  Gaussian $(\sigma_{\tau}=0.05)$ noise in \eqref{eq:noise}. 
The likelihood for dataset $\mathcal{D}$ conditioned on $v$ can be written as 
\begin{equation}
    p(\mathcal{D}|v)\propto\prod_{i=0}^{n_\tau}\frac{1}{\sqrt{2\pi\sigma_{\tau}^2}}\exp(-\frac{(T_d(x_{i})-\frac{x_i}{v})^2}{2\sigma_{\tau}^2T^{2}(x_{i})}).
\end{equation}
$T$ and $T_d$ are the ground truth and noisy measurements of the travel-time. 
To simplify the derivation, we consider finding $p(v^{-1}|\mathcal{D})$ instead of $p(v|\mathcal{D})$. 
We assume that the prior distribution $p(v^{-1})$ of $v^{-1}$ obeys Gaussian distribution $\mathcal{N}(0,\sigma^2_0)$. As a result, the posterior distribution could be calculated through
\begin{small}
\begin{equation}
\begin{aligned}
&p(v^{-1}|\mathcal{D})=\frac{p(\mathcal{D}|v^{-1})p(v^{-1})}{p(\mathcal{D})}\propto\\&
\prod_{i=0}^{n_\tau}\frac{1}{\sqrt{2\pi}{\sigma_{\tau}T(x_i)}}\exp(-\frac{(T_d(x_i)-x_iv^{-1})^2}{2\sigma_{\tau}^{2}T^{2}(x_i)})\frac{1}{\sqrt{2\pi}\sigma_0}\exp(-\frac{(v^{-1})^2}{2\sigma_0^2}).
\end{aligned}
\label{eq:P(v|D)}
\end{equation}
\end{small}
As $p(v^{-1}|\mathcal{D})$ is the product of a series of Gaussian distributions, the analytical solution of $p(v^{-1}|\mathcal{D})$ could be derived based on equation \eqref{eq:P(v|D)}, which also obeys Gaussian distribution. The analytical solution of the posterior distribution for $v^{-1}$ can be written as follows:
\begin{equation}
\begin{split}
    \mu_{post}&=\frac{a\sigma_{0}^{2}}{b\sigma_{0}^{2}\sigma_{\tau}^{2}+a},\\
    \sigma_{post}&=\frac{\sigma_{0}^{2}\sigma_{\tau}^{2}T^2_d(x_1)T^2_d(x_2)}{b\sigma_{0}^{2}+\sigma_{\tau}^{2}T^2_d(x_1)T^2_d(x_2)},
\label{eq:analytic}
\end{split}
\end{equation}
where
\begin{equation}
\begin{split}
    a&=x_{1}T_d(x_1)T^2_d(x_2)+x_{2}T_d(x_2)T^2_d(x_1),\\ b&=x^2_2T^2_d(x_1)+x^2_1T^2_d(x_2).
\label{eq:analytic-2}
\end{split}
\end{equation}
In our case, the standard deviation of the prior distribution is $\sigma_{0}=1$ and the level of relative Gaussian noise $\sigma_n$ is 0.05. Two measurement data are located at $x_1=1$ km and $x_2=2$ km. The true value of $v^{-1}$ is 0.5.

To test the performance of BPINNs-VI and BPINNs-SVGD for quantifying uncertainty in BPINNs, a trainable parameter is used to approximate $v^{-1}$. The travel time $T(x)$ is approximated by a fully-connected network with 2 hidden layers and 20 neurons for each layer with the swish activation function after each layer. For VI, we update the parameters for 5000 epochs, and 100 samples of posterior distribution are collected. As for SVGD, 30 particles are chosen and trained for 5000 epochs. 100 evenly spaced points $x\in[0,2]$ $\rm{km}$ are used as the residual points for both methods. 

\begin{figure}
	\centering
		\includegraphics[width=2.50in,height=2.0in]{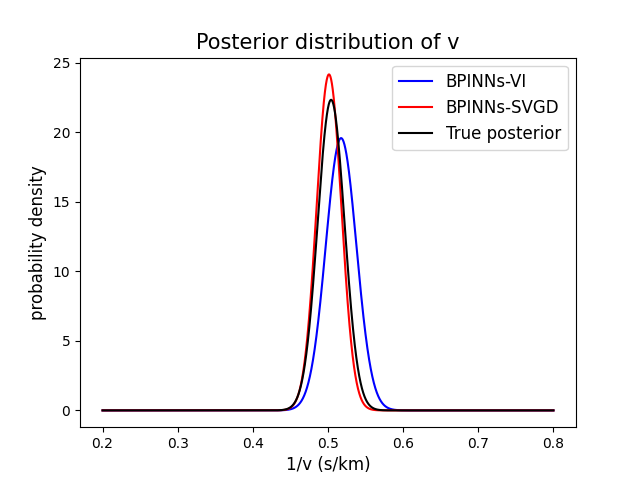}
	\caption{Posterior distributions for $v^{-1}$ with $5\%$ relative Gaussian noise by BPINNs-VI, BPINNs-SVGD and the true posterior distribution for the 1D homogeneous model. }
	\label{fig:v posterior}
\end{figure}

We compare the results from VI and SVGD with the analytic solution of $v^{-1}$ based on equation \eqref{eq:analytic}. From Table \ref{tab:example1}, we can see that the approximated mean and standard deviation of BPINNs-SVGD are closer to the true posterior mean and standard deviation than BPINNs-VI. Furthermore, Figure \ref{fig:v posterior} shows that BPINNs-SVGD with 30 particles fits better with the analytic distribution of $p(v^{-1}|\mathcal{D})$ than BPINNs-VI. It appears that BPINNs-SVGD is more accurate in estimating the posterior distribution of $v^{-1}$ in this example.

\begin{table}[h]
\centering
\caption{True and approximated posterior distribution of $v^{-1}$ for 1D homogeneous model with $5\%$ relative Gaussian noise.}
    \begin{tabular}{c c c}
      \hline
      Method &$\mu$&$\sigma$\\
      \hline
      True&0.5041 &0.0178\\
      VI& 0.5174 &0.0204\\
      SVGD&$\bm{0.5013}$ &$\bm{0.0165}$ \\
      \hline
    \end{tabular}
    \label{tab:example1}
\end{table}

\subsection{Cross-hole tomography}
\label{subsec: cross}

\begin{figure}[h]
\centering
\subfigbottomskip = 0pt
\subfigcapskip = -2pt
\subfigure[The ground truth]{\includegraphics[width = 0.24\textwidth]{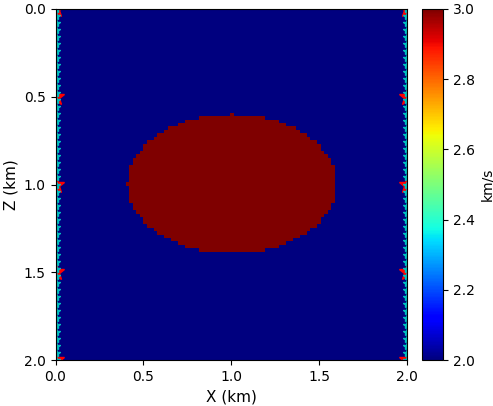}}\\
\subfigure[BPINNs-SVGD]{\includegraphics[width = 0.24\textwidth]{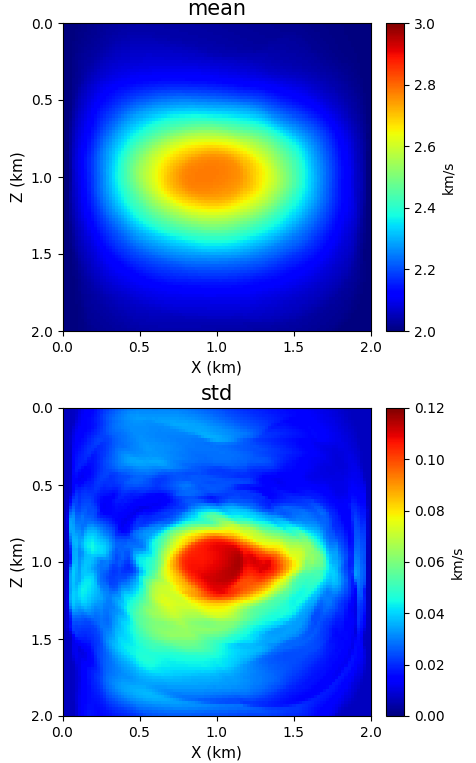}}
\subfigure[BPINNs-VI]{\includegraphics[width = 0.24\textwidth]{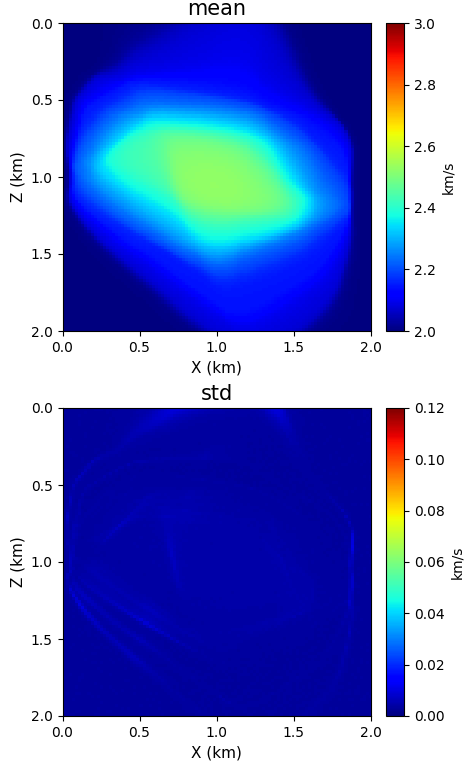}}
\caption{(a) The ground truth velocity model with an ellipsoidal anomaly for the cross-hole tomography, the predicted velocity model (mean) and approximated uncertainty ($\pm 1$ std) by (b) BPINNs-SVGD and (c) BPINNs-VI, respectively. Here, an ellipsoidal high-velocity anomaly of 3 km/s is embedded in a homogeneous background model (2 km/s). $2\times5$ sources with a uniform distance of 400 m are placed at $x = 0$ and 2 km (red star), $2\times51$ receivers and sample points of well logs are placed at $x = 0$ and 2 km with an interval of 40 m (cyan star).} 
\label{fig:inclusion} 
\end{figure}

Next, we consider a synthetic velocity model with ellipsoidal anomaly to investigate the performance of the BPINNs with VI and SVGD, shown in Figure \ref{fig:inclusion} (a). 
In this case,  5 equally spaced sources are placed on both left boundary ($x$ = 0 km), and the right boundary ($x$ = 2 km). 2 $\times$ 51 receivers are evenly spaced on the left boundary and right boundary. Two well-logs at $x$ = 0 and $2$ km provide 2 $\times$ 51 measured velocity data at the same locations as the receivers.
The $5\%$ Gaussian noise is added to the measurement data with $\sigma_{v}= \sigma_{\tau} =0.05$ in equation \eqref{eq:noise}.

For this problem, two independent fully connected neural networks are used to represent $\tau$ and $v$, respectively. For the travel-time factor, the neural network consists of 10 hidden layers with 20 neurons per layer with Swish activation function. For the velocity, we chose the network that includes 10 hidden layers with 10 neurons for each layer with ELU activation function. A sigmoid function is used to normalize the network output and scaling factors $v_{\rm{max}}$ and $v_{\rm{min}}$ introduced in \eqref{eq:normalize} are used to approximate predicted velocity. For SVGD, the prior distributions of $\Sigma_{\mathcal{D}_v}$, $\Sigma_{\mathcal{D_\tau}}$ and $\Sigma_{\mathcal{R}}$ are Gamma distribution $Gamma(2,10^{-6})$. In this example, the particle number of SVGD has been set to 5. The BPINNs with VI and SVGD are trained for 1000 epochs by Adam optimizer.
These configurations will be used in subsequent experiments unless otherwise specified.

Figure \ref{fig:inclusion} shows the results of BPINNs with VI and SVGD, respectively. Despite the fact that no direct measurements have been given in the anomaly area between 0.4 to 1.6 km, BPINNs with SVGD are capable of reconstructing the anomaly with the help of physics constraints. Due to the lack of information about the anomaly area, the velocity value of the anomaly is underestimated. Nonetheless, BPINNs indeed suggest higher uncertainty  around the anomaly. 
Furthermore, as the seismometers are located on the left and right boundaries, the associated uncertainty is lower than we expected. While SVGD provides a more accurate inversion of the velocity field as well as reasonable uncertainty estimation,  BPINNs with VI provide a less accurate estimation of the velocity field  and appear to be overconfident in the predicted results.

\begin{figure}[h] 
\centering
\subfigure[The ground truth travel-time]{\includegraphics[width = 0.23\textwidth]{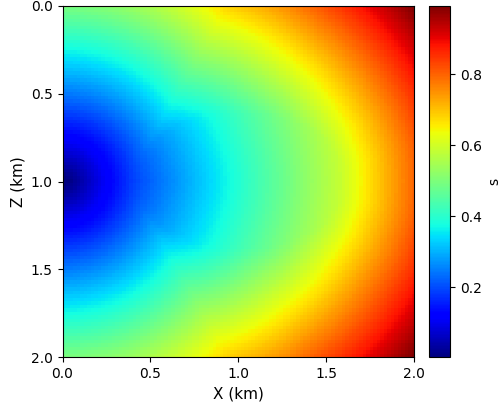}}\\
\centering
\subfigure[The predicted travel-time by SVGD]{\includegraphics[width = 0.23\textwidth]{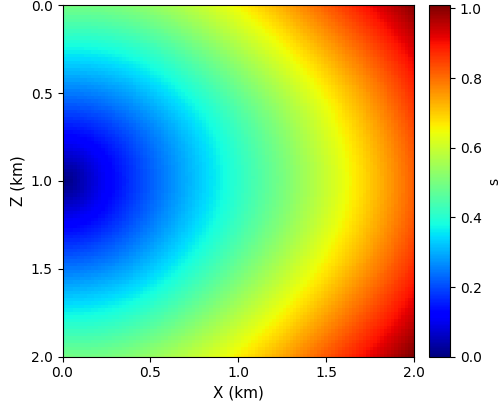}} \hspace{0.5cm}
\subfigure[Contours for SVGD and ground truth]{\includegraphics[width = 0.19\textwidth]{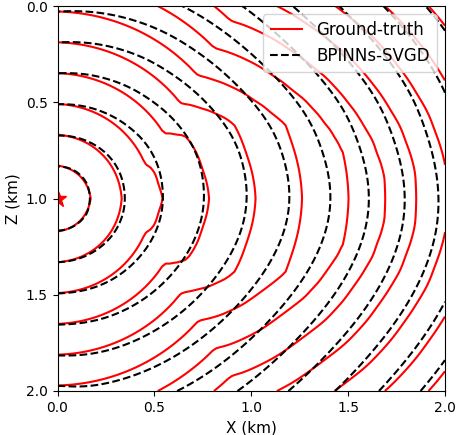}}\\
\subfigure[The predicted travel-time by VI]{\includegraphics[width = 0.23\textwidth]{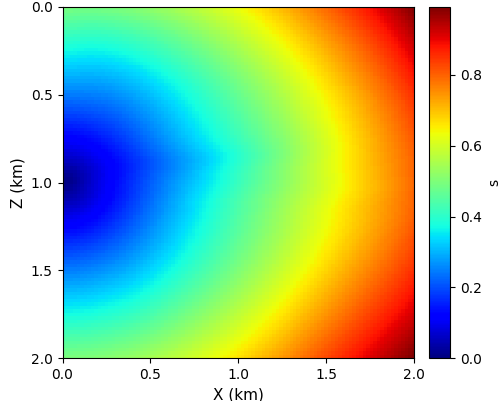}} \hspace{0.5cm}
\subfigure[Contours for VI and ground truth]{\includegraphics[width = 0.19\textwidth]{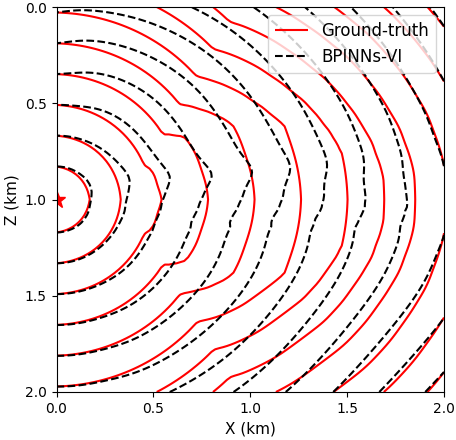}}
\caption{(a) The ground truth travel-time field computed by Fast Marching Method with sckit-fmm toolkit when the source is located at (0 km, 1 km) for the model with an ellipsoidal anomaly. The predicted travel-time field by (b) BPINNs-SVGD and (d) BPINNs-VI with $5\%$ Gaussian noise. The travel-time contours of the ground truth and predicted by (c) BPINNs-SVGD and (e) BPINNs-VI.}
\label{fig:time_inc}
\end{figure}

To further demonstrate the performance of BPINNs, we compared the ground truth and predicted travel-time field by BPINNs with SVGD and VI in Figure \ref{fig:time_inc}. Compared with results by BPINNs-VI, BPINNs-SVGD provided a more accurate travel-time field. Nonetheless, the predicted travel-time fields appear to be smoother than the ground truth one, and BPINNs cannot capture the detailed effects induced by the inclusion. Besides, Table \ref{tab:2d} lists the quantitative comparison of absolute relative error (ARE) and correlation coefficient $\gamma$
of BPINNs with VI and SVGD, indicating that the BPINNs-SVGD performs better for the cross-hole tomography in both performance metrics.

\begin{table}[h]
\caption{The error metric for inverted velocity field ($v$) and reconstructed traveltime ($T$) by BPINNs with VI and SVGD for the cross-hole tomography.}
\centering
    \begin{tabular}{c c c c c}
      \hline
      BPINN&
      ARE ($v$)& 
      $\gamma$ ($v$)&ARE ($T$)& $\gamma$ ($T$) \\
      \hline
      VI&
      0.0803 &
      0.7889&0.0450&0.9925 \\
      SVGD& \bf{0.0748} & \bf{0.8513}& \bf{0.0380} & \bf{0.9957}\\
      \hline
    \end{tabular}
    \label{tab:2d}
\end{table}

\subsection{Surface tomography}
In this subsection, we shall investigate the influence of different levels of noise on surface tomography by BPINNs-SVGD. For surface tomography, the travel-time data are often collected on the surface, therefore less information about the deeper layer can be provided, resulting  in the uncertainty increase with depth. With this prior knowledge, depth-dependent uncertainty is introduced to better characterize the uncertainty.

\begin{figure}[h]
	\centering
		\includegraphics[scale=.80]{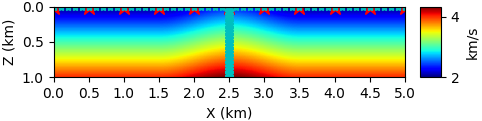}
	\caption{The ground truth velocity model. 11 sources are equispaced at $z=0$ km (red star). 51 receivers are uniformly spaced at $z =0$ km. 50 equispaced receivers with velocity sample points are located at $x=2.5$ km (cyan star).}
	\label{fig:arch}
\end{figure}

Figure \ref{fig:arch} shows the ground truth velocity for the surface tomography. At the surface of the model ($z$ = 0 km), 11 evenly-spaced sources with a distance of 0.5 km between each source location are placed and 51 evenly-spaced receivers  with a distance of 100 m between each sensor location. In addition,  50 even-spaced receivers along $x$ = 2.5 km with a distance of  20 m between each sensor location. A well log with 50 evenly-spaced receivers across $x$ = 2.5 km is used to measure the velocity. The measured travel-time and velocity are perturbed by $5\%$, $15\%$, and $25\%$ three different levels of relative Gaussian noise. The architecture and hyperparameters of BPINNs-SVGD are the same as the section \ref{subsec: cross}.

\begin{figure}[h]
\centering
\subfigure[Depth independent uncertainty]{\includegraphics[width = 0.40\textwidth]{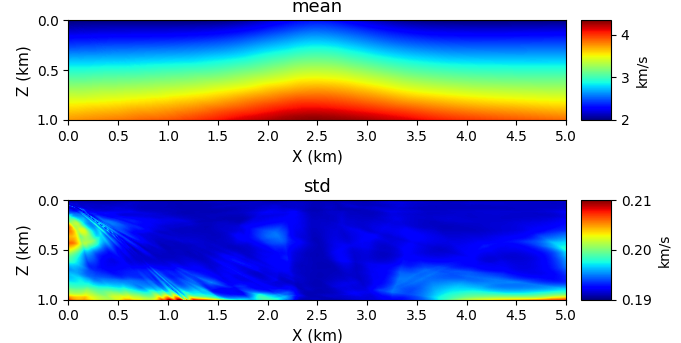}}
\subfigure[Depth dependent uncertainty]{\includegraphics[width = 0.40\textwidth]{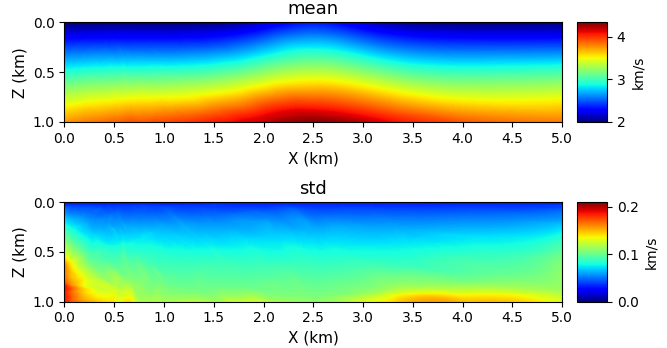}}
\caption{ The predicted mean and standard deviation based on BPINNs-SVGD  with (a) depth-independent uncertainty and (b) depth-dependent uncertainty under $5\%$ Gaussian noise. }
\label{fig:compare depth depend} 
\end{figure}

The predicted velocity and the uncertainty estimates by BPINNs-SVGD with $5\%$ Gaussian noise are shown in Figure \ref{fig:compare depth depend} (a). BPINNs-SVGD appears to overestimate the uncertainty at the shallow depth while underestimating it as the depth increases. The uncertainty shown in Figure \ref{fig:compare depth depend} (a) tends to  be more uniform over the domain due to the depth-independent data uncertainty introduced in equation \eqref{eq:uncertainty svgd}. 
To address this issue, we introduced the depth-dependent uncertainty with depth discussed in section \ref{bnn-ek}. The prior distribution of $\Sigma_{\mathcal{D}_v}(z_{\min})$ and $\Sigma_{\mathcal{D}_v}(z_{\max})$
are represented by Gamma distributions: $Gamma(\Sigma_{\mathcal{D}_v}(z_{\min})|2,10^{-6})$ and $Gamma(\Sigma_{\mathcal{D}_v}(z_{\max})|1.5,10^{-6})$, respectively.
According to our experiments, the results are insensitive to the parameters of the distribution when they are restricted in a reasonable range. 
Figure \ref{fig:compare depth depend} (b) shows the results by BPINNs-SVGD with the uncertainty with respect to the depth. After adopting depth-dependent uncertainty, the predicted uncertainty provides a much more reasonable uncertainty in the spatial domain, especially with respect to the depth $z$.

\begin{figure}[h]
\centering
\subfigure[Depth independent uncertainty]{\includegraphics[width = 0.40\textwidth]{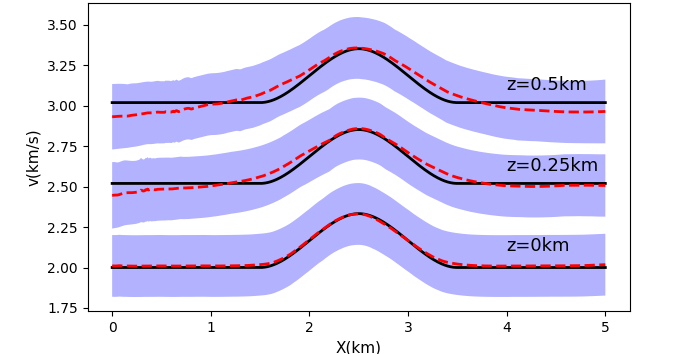}}
\subfigure[Depth dependent uncertainty]{\includegraphics[width = 0.40\textwidth]{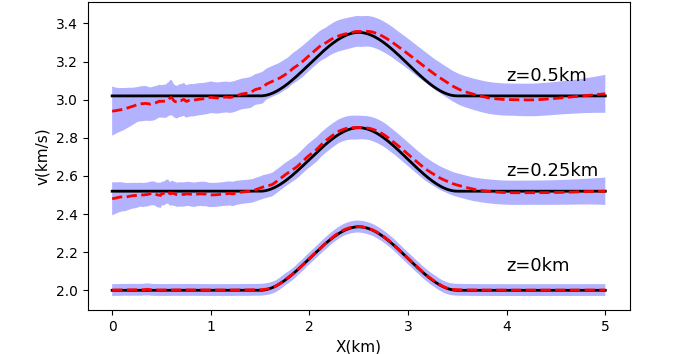}}
\caption{The predicted velocity model and corresponding uncertainty under $5\%$ noise with/without assumption of depth-dependent uncertainty. The black solid line represents the ground truth velocity, red dotted line represents the predicted velocity. The shaded blue areas are uncertainty covered by $\pm 1$ std.} 
\label{fig:z00.51} 
\end{figure}

\begin{figure}[h] 
\centering
\subfigure[The ground truth travel-time]{\includegraphics[width = 0.40\textwidth]{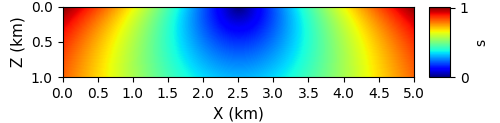}}
\subfigure[The predicted travel-time]{\includegraphics[width = 0.40\textwidth]{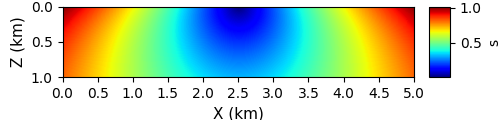}}
\subfigure[Contour of the predicted and true travel-time]{\includegraphics[width = 0.40\textwidth]{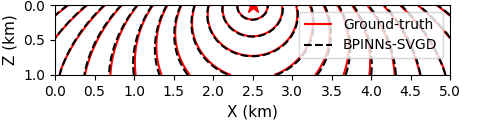}}
\caption{(a) The ground truth travel-time field computed by Fast Marching Method with sckit-fmm toolkit. (b) The predicted travel-time field by BPINNs-SVGD with $5\%$ Gaussian noise. Here, the source is located at (2.5 km, 0 km). (c) The contour of predicted and ground truth travel-time.}
\label{fig:compare time}
\end{figure}

Figure \ref{fig:z00.51} compares the predicted velocity and the uncertainty based on SVGD with/without depth-dependent uncertainty at  depth  $z$ = 0, 0.25 and 0.5 km, respectively. For instance, 51 receivers are placed at $z=0$ and the information is rich enough for BPINNs-SVGD to precisely predict the velocity with lower uncertainty. As the depth increases, the data provides less information, resulting in a larger deviation from the ground truth velocity and higher uncertainty can be expected. By introducing the depth-dependent uncertainty, the overestimation of uncertainty at shallow depth shown in SVGD with depth-independent uncertainty (Fig \ref{fig:z00.51} (a)) can be significantly reduced as in Figure \ref{fig:z00.51} (b).
In addition, the predicted travel-time field by BPINNs-SVGD with depth-dependent uncertainty is displayed in Figure \ref{fig:compare time}. It can be seen that the predicted travel-time matches well the ground truth, which demonstrates the quality of the forward approximation of BPINNs.

\begin{figure}[h]
\centering
\subfigure[$15\%$ relative noise]{\includegraphics[width = 0.40\textwidth]{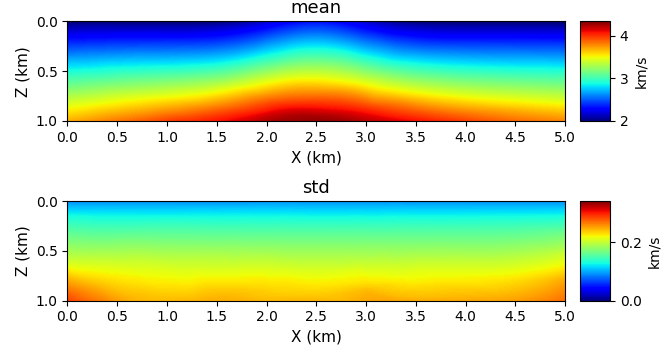}}
\subfigure[$25\%$ relative noise]{\includegraphics[width = 0.40\textwidth]{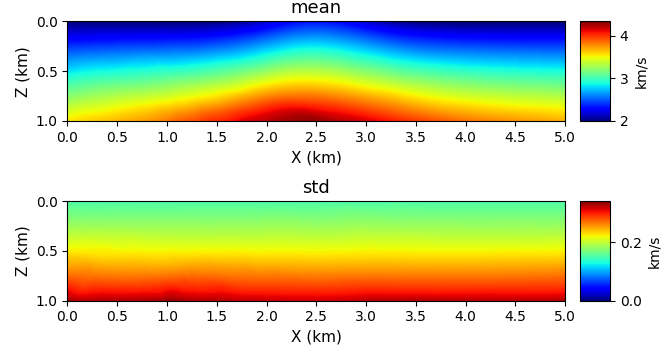}}
\caption{Predicted results by the BPINNs-SVGD  under (a) $15\%$ relative Gaussian noise and (b)  $25\%$ relative noise. The uncertainty of predicted results are represented by $\pm 1$  std.}\label{fig: noise-0.15-0.25} 
\end{figure}

To further study the noise influence on the performance of BPINNs, we test the proposed method over the observed data with $15\%$ and $25\%$ Gaussian noise based on the  depth-dependent velocity uncertainty $\Sigma_{\mathcal{D}_{v}}(z)$ in \eqref{eq:depth uncertainty} by BPINNs-SVGD. The predicted velocity model and uncertainty are shown in Figure \ref{fig: noise-0.15-0.25}. Even with $25\%$ Gaussian noise, BPINNs can still accurately recover the velocity model, demonstrating the robustness to noise of BPINNs. The corresponding error metrics are shown in Table \ref{tab: detail}. As expected, the ARE and estimated uncertainty increase with the noise level.  With $25\%$ noise included in the measured data, the ARE is less than $2\%$ and correlation reaches $98\%$, demonstrating the robustness of BPINNs. In addition,  as we expected,  the predicted uncertainty by BPINNs increased as the depth increases as shown in Figure \ref{fig: noise-0.15-0.25}. 

\begin{table}[tbp]
\caption{Error metric of inverted velocity by BPINN-SVGD for surface tomography with $5\%, 15\%, 25\%$ Gaussian noise.}
    \begin{tabular}{c c c c}
      \hline
      Level of noise&
      ARE& 
      $\gamma$&
      Uncertainty(mean/max std)\\
      \hline
      0.05&
      0.0107&
      0.9972&
      (0.0877/0.1895)\\
      0.15&0.0145&0.9938&(0.1868/0.2865)\\
      0.25&0.0196&0.9885&(0.2313/0.3338)\\
      \hline
    \end{tabular}
    \label{tab: detail}
\end{table}

\section{Discussion}
\label{sec:discussion}
In the previous section, we tested BPINNs on two relatively simple benchmarks. Here, we shall discuss the performance of the proposed scheme on a more complex and realistic model. The overthrust model is a 3-D geological model proposed as a result of collaboration by over fifty organizations. It is built with erosional truncation and sediment that cover the basement blocks. The overthrust model includes complex structures like reverse faults and converging thrusts. Because of its 
complicated geological characteristics, the overthrust model is used to test different imaging and inversion algorithms \cite{aminzadeh19963}.

\begin{figure}[h]
\centering
\subfigure[Ground truth velocity model]{\includegraphics[width = 0.33\textwidth]{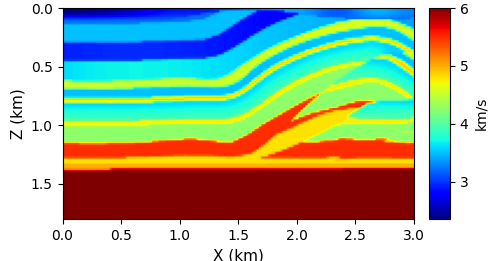}}\\
\subfigure[Predicted velocity model]{\includegraphics[width = 0.33\textwidth]{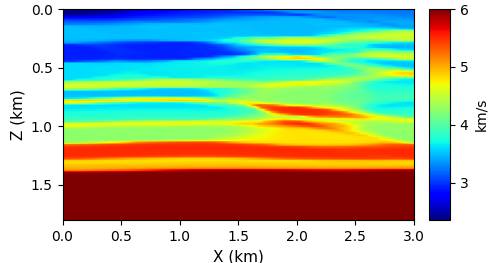}}
\subfigure[Uncertainty about the predicted velocity model]{\includegraphics[width = 0.33\textwidth]{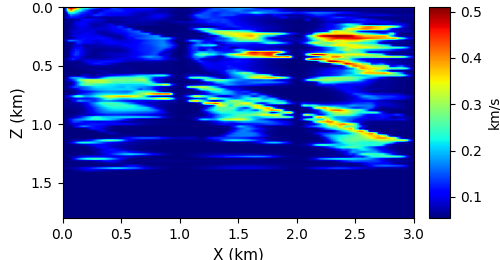}}
\caption{(a) The ground truth velocity of the overthrust model. The size of model is $91\times151$ grids with a grid size of 20 m. (b) The predicted velocity model of BPINNs-SVGD. (c) The estimated uncertainty displayed by $\pm 1$ std of predicted overthrust model.}
\label{fig:overthrust true_pred} 
\end{figure}

We have intercepted a 2D slice from the 3D overthrust model as shown in Figure \ref{fig:overthrust true_pred} (a). To acquire the observed travel-time for training, $11\times2$ evenly-paced sources located on the left boundary ($x = 0$ km) and right boundary ($x = 3$ km) are chosen, while two series of receivers located on the left and right boundaries with an interval of 0.02 km provide $2\times91$ travel-time data for each source. Four well logs located at $x=0,1,2,3$ km provide velocity data with an interval of 0.02 km along the depth. Given its complex structure, the network for inferring the velocity has been set to 10 hidden layers with 40 neurons and noise-free training data are used.
Other Settings are the same as section \ref{subsec: cross}.

The estimated velocity model by BPINNs-SVGD is shown in Figure \ref{fig:overthrust true_pred} (b). Even though the details like the thrusts are not accurately captured by the proposed method, it does capture well the relatively simple structure on the left half domain of the overthrust model. Furthermore, the location of the high-speed region in the deeper layer is well captured. 
We also show the estimated uncertainty about the predicted velocity by BPINNs-SVGD for the overthrust model in Figure \ref{fig:overthrust true_pred} (c). Since we can access the velocity data at $x=0, 1, 2, 3$ km, the estimated uncertainty is relatively low at these locations. Because of its simple structure, the prediction in the homogeneous region at depth ranging from 1.4 to 1.9 km presents a lower uncertainty. For the complex arch and fault structure in the overthrust model from $x=1.5$ to 3 km, higher uncertainty is predicted as expected, indicating the effectiveness of the uncertainty estimated by BPINNs.

\begin{figure}[h]
\subfigure[The ground truth travel-time]{\includegraphics[width = 0.30\textwidth]{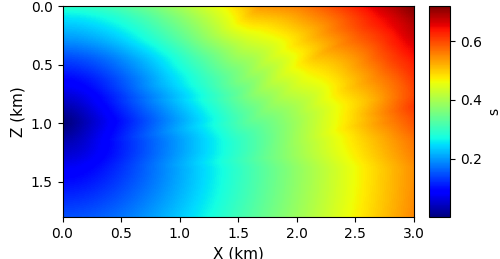}}
\subfigure[The predicted travel-time]{\includegraphics[width = 0.30\textwidth]{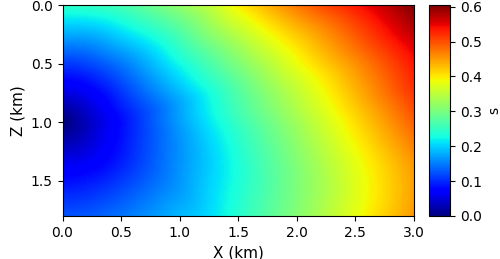}}
\centering\subfigure[Contour of the predicted and ground truth travel-time]{\includegraphics[width = 0.26\textwidth]{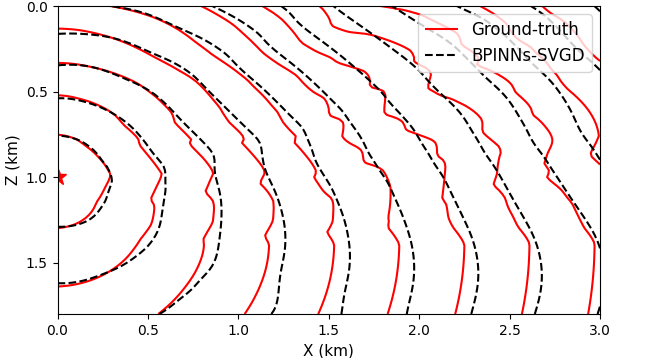}}
\caption{(a) Ground truth travel-time, (b) predicted travel-time, (c) contour of the predicted and true travel-time for the overthrust model when source-point is at (0km, 1km).}
\label{fig:overthrust traveltime} 
\end{figure}

Figure \ref{fig:overthrust traveltime} shows the ground truth and predicted travel-time by BPINNs-SVGD for the overthrust model. Even though with some details are missing, the  BPINNs-SVGD predictions can capture the rough travel-time field.

Given that the eikonal equation describes the relationship between the first-arrival time and the velocity, it contains significantly less information than the full waveform inversion. 
Besides, overthrust is a relatively complex geological model. It is challenging for travel-time tomography to perfectly invert such a velocity model. Nonetheless, the predicted velocity model  from  BPINNs-SVGD could  be an effective initial model for FWI or migration imaging.  In addition, the uncertainty quantification by the proposed method can provide a meaningful assessment of  the reliability of the predictions.

\section{Conclusion}
\label{sec:conclusion}
We introduced the BPINNs for seismic travel-time tomography based on the eikonal equation. For the Gaussian VI and SVGD algorithms are adopted to infer the velocity and the corresponding uncertainty quantification. To better account for the higher uncertainty that exists in the deep layer for surface tomography, we further introduced depth-dependent uncertainty as the prior knowledge. After demonstrating the effectiveness of BPINNs-SVGD in 1D homogeneous model and ellipsoidal model, we adopt it to study the influence of different levels of noise on predictive uncertainty for surface tomography. We also employ BPINNs-SVGD for a more realistic  overthrust model. Despite its complex structure and insufficient information, the proposed method provides a reasonably accurate velocity model and meaningful uncertainty estimation, demonstrating the feasibility and potential of the proposed method for realistic applications. If a better velocity model is needed, the suggested method  
may serve  as an effective initial model for FWI.


%




\ifCLASSOPTIONcaptionsoff
  \newpage
\fi



%
\bibliographystyle{IEEEtran}
\bibliography{tgrs-ref.bib}

%

\begin{IEEEbiography}[{\includegraphics[width=1in,height=1.25in,clip,keepaspectratio]{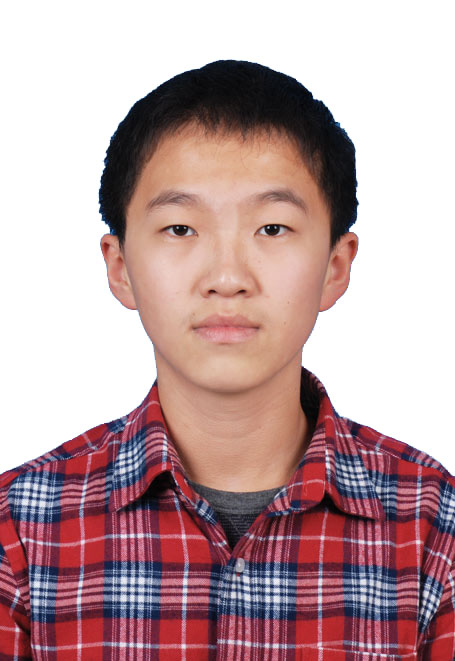}}]{Rongxi Gou} received the B.S. degree in Information engineering in 2021 from Xi'an Jiaotong University. Now, he is studying for a master's degree in Information and Communications Engineering at Xi 'an Jiaotong University. His research interests mainly include machine learning for forward and inverse problems, optimization theory and signal processing.
\end{IEEEbiography}

\begin{IEEEbiography}[{\includegraphics[width=1in,height=1.25in,clip,keepaspectratio]{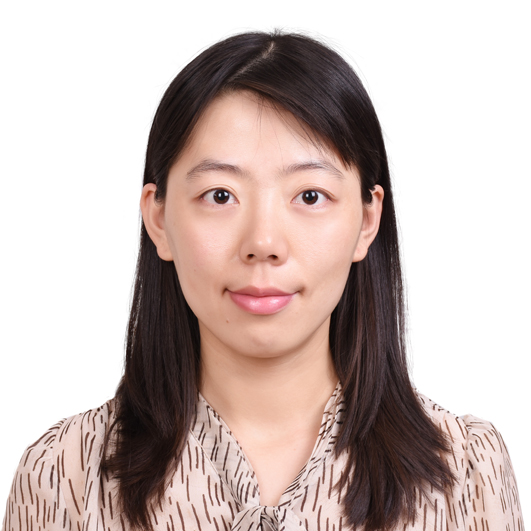}}]{Yijie Zhang} received the B.S. degree in Information engineering in 2010 from Xi'an Jiaotong University, Shaanxi, China, and the Ph.D. degree in Information and Communication engineering in 2016 from Xi'an Jiaotong University, Shaanxi, China. She was a visiting scholar with Department of Mathematics, the University of Iowa. She is currently an associate professor with School of Information and Communications Engineering, Xi'an Jiaotong University. Her research interests include machine learning in wave-fields forward modeling and inversion, signal processing; wave propagation theory and numerical simulation in complex media.
\end{IEEEbiography}

\begin{IEEEbiography}[{\includegraphics[width=1in,height=1.25in,clip,keepaspectratio]{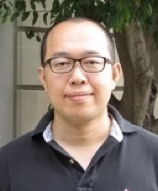}}]{Xueyu Zhu} received his Ph. D. degree in applied mathematics in 2013 from Brown University, Providence, RI. He is currently an associate professor at the Department of Mathematics, University of Iowa. He is also affiliated with an interdisciplinary Ph. D. program in Applied Mathematical and Computational Sciences. His research interests lie in computational mathematics, scientific computing, uncertainty quantification, model reduction, and machine learning.
\end{IEEEbiography}

\begin{IEEEbiography}[{\includegraphics[width=1in,height=1.25in,clip,keepaspectratio]{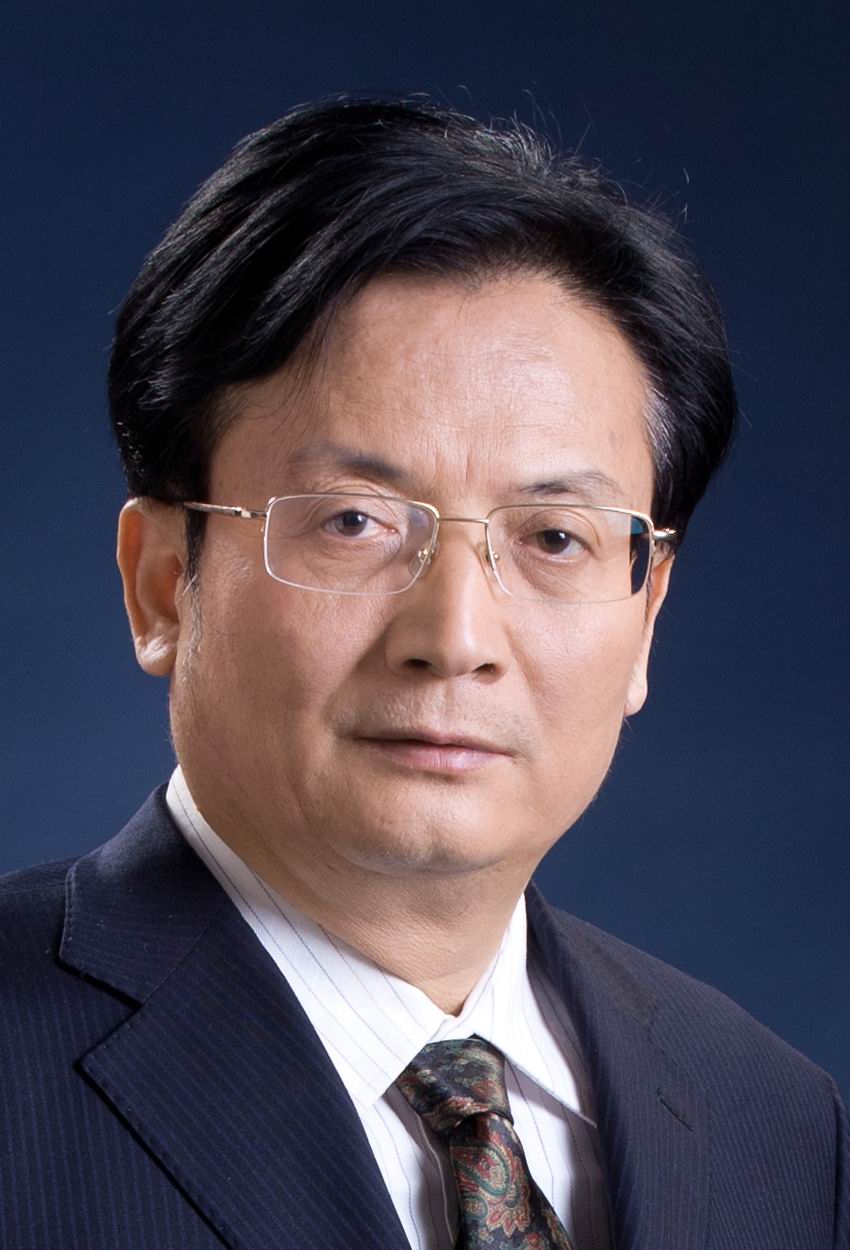}}]{Jinghuai Gao} received the M.S. degree in applied geophysics from Chang’an University, Xi’an, China, in 1991 and the Ph.D. degree in electromagnetic field and microwave technology from Xi’an Jiaotong University, Xi’an, in 1997. From 1997 to 2000, he was a Postdoctoral with the Institute of Geology and Geophysics, Chinese Academy of Sciences, Beijing, China. In 1999, he was a Visiting Scientist with the Modeling and Imaging Laboratory, University of California, Santa Cruz, CA, USA. He is currently a Professor with the School of Information and Communications Engineering and the School of Mathematics and Statistics, Xi’an Jiaotong University. His research interests include seismic wave propagation and imaging theory, seismic reservoir and fluid identification, and seismic inverse problem theory and methods.
\end{IEEEbiography}




\end{document}